\documentclass[aps,prl,twocolumn,longbibliography]{revtex4-1}

\usepackage{graphics, bm, xspace}
\usepackage{graphicx}
\usepackage{psfrag}
\usepackage{amsmath}
\usepackage{amssymb}
\usepackage{epsfig}
\usepackage{subfigure}
\usepackage{grffile}
\usepackage{times}
\usepackage{color}
\usepackage{multirow}
\usepackage[bookmarks=false,linkcolor=blue,urlcolor=blue,colorlinks,citecolor=blue]{hyperref}
\usepackage{float}
\usepackage{gensymb}

\newcommand{\beq}    {\begin{equation}}
\newcommand{\enq}    {\end{equation}}
\newcommand{\ceq}[1] {(\ref{#1})}

\newcommand{\kk}     {{\bf k}}

\newcommand{\up} {\uparrow} 
\newcommand{\down} {\downarrow}

\usepackage{amssymb}
\usepackage{amsbsy}
\usepackage{amsmath}
\usepackage{graphicx}
\usepackage{graphics}
\usepackage{setspace}
\usepackage{array}
\usepackage{color}
\usepackage{fontenc}
\usepackage{textcomp}
\usepackage{bm}
\usepackage{float}
\usepackage[bookmarks=false,linkcolor=blue,urlcolor=blue,colorlinks,citecolor=blue]{hyperref}
\usepackage{hyperref}

\newcommand{\vex}[1]{\bm{\mathrm{#1}}}

\newcommand{\bsub}{\begin{subequations}}
\newcommand{\esub}{\end{subequations}}

\newcommand{\hsm}{h_{\rm SM}}
\newcommand{\hsc}{h_{\rm SC}}
\newcommand{\mx}{\mathcal{M}_x}
\newcommand{\my}{\mathcal{M}_y}

\graphicspath{{../Figures/}}

\begin{document}
\title{Second-order Dirac superconductors and magnetic field induced Majorana hinge modes}
\author{Sayed Ali Akbar Ghorashi$^1$, Xiang Hu$^1$, Taylor L. Hughes$^2$, Enrico Rossi$^1$}
\affiliation{$^1$Department of Physics, William $\&$ Mary, Williamsburg, Virginia 23187, USA}
\affiliation{$^2$Department of Physics and Institute for Condensed Matter Theory,
University of Illinois at Urbana-Champaign, IL 61801, USA}


\newcommand{\be}{\begin{equation}}
\newcommand{\ee}{\end{equation}}
\newcommand{\bea}{\begin{eqnarray}}
\newcommand{\eea}{\end{eqnarray}}
\newcommand{\h}{\hspace{0.30 cm}}
\newcommand{\vs}{\vspace{0.30 cm}}
\newcommand{\n}{\nonumber}
\begin{abstract}
We identify three dimensional higher-order superconductors
characterized by the coexistence of one-dimensional Majorana hinge states
and gapless surface sates.
We show how such superconductors can be obtained starting from the model of a spinful quadrupolar semimetal with two orbitals
and adding an s-wave superconducting pairing term. By considering all the possible s-wave pairings satisfying Fermi-Dirac statistics we obtain six different superconducting models.
We find that for two of these models a flat-band of hinge Majorana states
coexist with surface states, and that
these models have a non-vanishing quadrupole-like topological invariant.
Two of the other models, in the presence of a Zeeman term,
exhibit helical and dispersive hinge states
localized only at two of the four hinges. We find that these states are protected by combinations of rotation and mirror symmetries, and that the pair of corners exhibiting hinge states switches upon changing the
sign of the Zeeman term. Furthermore, we show that these states can be localized to a single hinge with suitable perturbations.
The remaining two models retain gapless bulk and surface states that spectroscopically obscure any possible hinge states.

\end{abstract}
\maketitle


The modern theory of polarization for crystalline insulators
\cite{Resta1994}
has revealed that in crystals a dipole moment can
be expressed in terms of Berry phases,
and that a finite dipole necessarily implies the presence of boundary charges.
The presence of nontrivial Berry phases and boundary states are the hallmarks of topological
systems~\cite{Chiu2016}, and indeed it is now clear that there is a strong connection between the
theory of topological insulators (TIs) and systems with quantized dipole moments\cite{zak1988,hughes2011,turner2012}.
This connection has led to the realization that the extension of the modern theory of polarization to higher multipole moments
allows the identification of new classes of topological crystalline insulators~\cite{Benalcazar2017-1,Benalcazar2017-2}, termed
``higher-order'' TIs (HOTIs). Within this framework, a higher-order multipole TI of order $m$ has a quantized
nonzero electric $m^{\rm th}$-pole in the bulk (with $m=1$ for a dipole, $m=2$ for a quadrupole,...) and localized charges
at its $d-m$-dimensional boundaries, $d$ being the insulator's spatial dimension.
Since the work of Refs. \onlinecite{Benalcazar2017-1,Benalcazar2017-2} many proposals of HOTIs of various types
have been presented~\cite{Schindler2018-1,Song2017,Schindler2018-2,EzawaPRL2018, EzawaPRB2018}.
Higher-order topological insulating phases have been realized
in metamaterial arrays~\cite{Noh2018, Peterson2018, Imhof2018}, phononic systems \cite{Serra-Garcia2018}, and it has been proposed that bismuth~\cite{Schindler2018-2}, strained SnTe\cite{Schindler2018-1}, and some 2D transition metal dichalcogenides~\cite{Ezawa}
are second-order TIs.
In addition, there has been exciting new work on higher order topological superconductors (HOTSCs) and topological semimetals (HOTSMs) ~\cite{Langbehn2017,Song2017,Wang2018-TH,Khalaf2051,CAlugAru2018,Ezawa,Wang2018-BA,wieder2018,wieder2019,Dassarma2018,FanZhang2018,ZhongWang2018,Loss2018,Loss2019,ChuanweiZhang19}.
\begin{figure}[htb]
\includegraphics[width=0.37\textwidth]{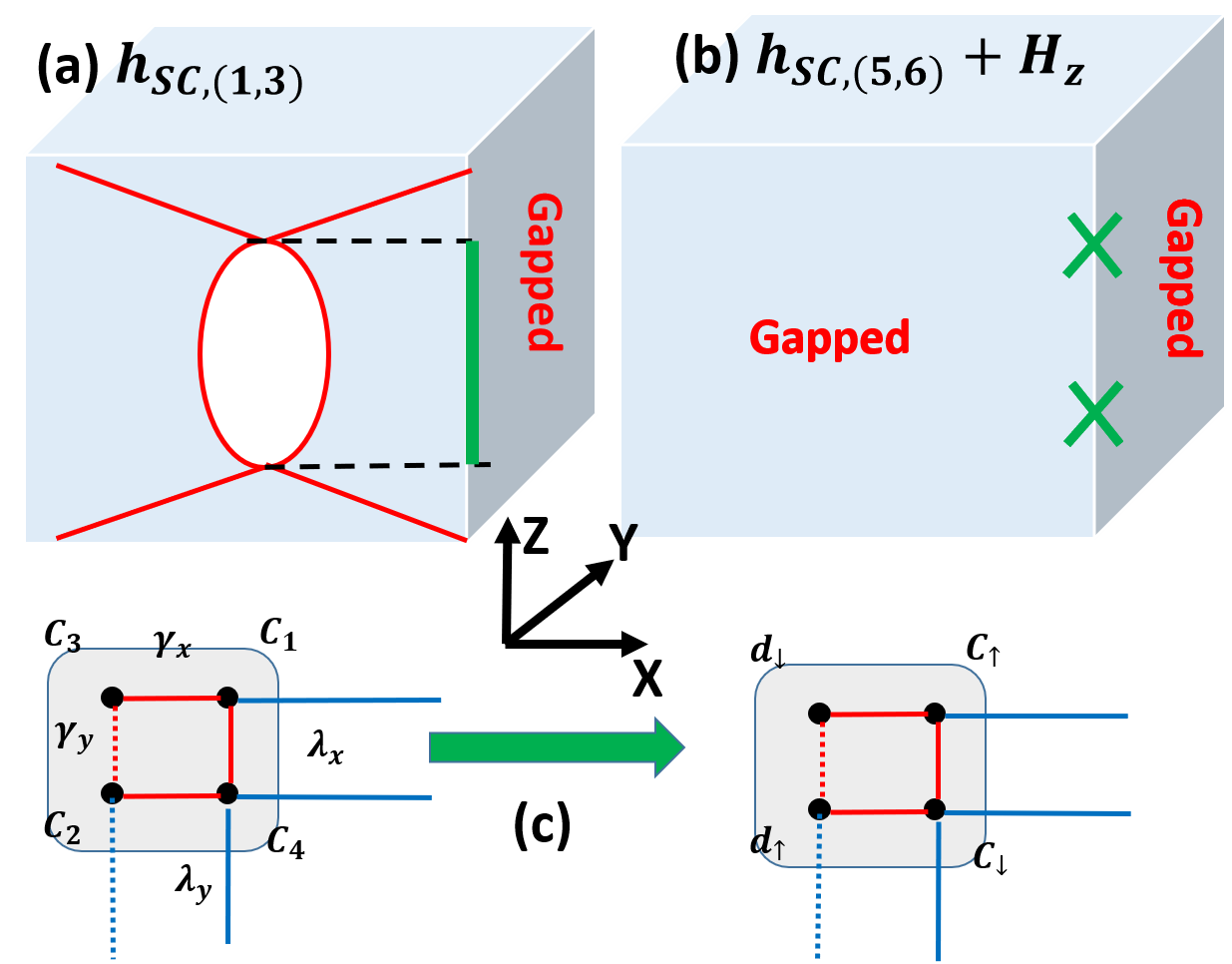}
\caption{Summary illustration of the models with hinge modes studied in this work. (a) model $h_{\rm SC,1}$ (rotate by $C_4$ for model $h_{\rm SC,3}$). (b) models $h_{\rm SC,5}$, $h_{\rm SC,6}$ with magnetic field (Zeeman term $H_z$) in the $z$-direction. Red lines represent surface states for $k_{x,y}=0$ cuts. Green lines represent flat-band hinge states; green crosses represent dispersing hinge nodes. (c) The unit cell convention used to convert between spinless and spinfull version of HOTI and HOTSM.
 }
\label{adfig}
\end{figure}

In this work we identify for the first time a new class of higher-order topological superconductors: {\em second-order Dirac superconductors}, Fig.~\ref{adfig}. Such novel three-dimensional (3D) higher-order topological superconductors exhibit the unique property to have both topologically protected Dirac cones for states at the surface (d-1 boundary) and topologically protected dispersionless Majorana states at the hinges (d-2 boundary). The hinge states can be interpreted as Majorana arcs joining the surface Dirac points. We explicitly show that this unique configuration of surface and hinge states is associated with a topological invariant that can be interpreted as the generalization to the superconducting case of the quadrupole moment.
This model can provide a platform to study the interplay of the bulk, surface and hinge states on an equal footing which can significantly help towards understanding the transport properties of systems with higher-order topology.
In addition, we identify 3D superconducting systems that can be driven into a higher-order topological state, exhibiting dispersive helical Majorana states at the hinges, simply via the application of an external magnetic field. Interestingly, we find that the hinges are localized at only two of the four
hinges (see \cite{wieder2018} and \cite{benalcazar2019} for insulating examples of states localized at two corners). Furthermore, the pairs of corners where the Majorana modes are located can be selected by changing the sign of the magnetic field. We also show that by breaking diagonal/antidiagonal mirror symmetries, it is possible to drive the system into a unique topological state in which Majorana states are present only at one of the hinges.

We start by considering a model for a topological quadrupolar semimetal constructed from layers of 2D quadrupolar topological insulators (TIs) \cite{Lin2017}. Schematically, the unit cell for the tight binding model for the 2D quadrupolar insulator layer is illustrated in Fig.~\ref{adfig}~(c).
For each cell we have two orbitals ($c$, $d$) and a spin-1/2 degree of freedom, represented by the four black dots in Fig.~\ref{adfig}~(c).
Let $\gamma_i$ ($i=x,y$) be the intra-cell hopping amplitudes, red lines in Fig.~\ref{adfig}~(c),
and $\lambda_i$ the inter-cell hopping, blue lines in Fig.~\ref{adfig}~(c).
Hopping processes represented by dotted lines have a phase that is opposite to the one of the hopping processes
represented by solid lines.
Depending on the choice of interlayer tunneling terms between 2D layers of quadrupolar TIs
we obtain different Hamiltonians, $H$, for the resulting 3D system.
In momentum space we have $H_{\rm SM}=\sum_{\kk}\psi_{\kk}^\dagger \hsm(\kk) \psi_{\kk},$ where
$\psi_\kk^T$ is the spinor
$(c_{\kk\up},d_{\kk\up},d_{\kk\down},c_{\kk\down})$, formed by the annihilation operators
$c_{\kk\alpha}$, $d_{\kk\alpha}$,
for an electron in orbital $c$, $d$, with spin $\alpha$ and momentum $\kk$,
and $\hsm$ is a $4\times 4$ Bloch Hamiltonian matrix of the general form:
\begin{align}
 \hsm(\vex{k})=(\gamma_x+\chi_x(k_z)+\lambda_x\cos(k_x))\Gamma_4+\lambda_x\sin(k_x)\Gamma_3\cr
 +(\gamma_y+\chi_y(k_z)+\lambda_y\cos(k_y))\Gamma_2+\lambda_y\sin(k_y)\Gamma_1.
 \label{eq:Ham}
\end{align}
In Eq.~\ceq{eq:Ham} all the lattice constants are taken to be 1, $\chi_i(k_z)$ ($i=x,y$) are periodic functions of $k_z$ with forms fixed by the interlayer tunneling terms,
$\{\Gamma_\alpha\}$ are the $4\times4$ matrices given by the direct product of
$2\times2$ Pauli matrices $\sigma_i$, $\kappa_i$ in spin and orbital space, respectively:
$\Gamma_0=\sigma_3\kappa_0$,
$\Gamma_i=-\sigma_2\kappa_i$,
$\Gamma_4=\sigma_1\kappa_0$.
In the remainder
we assume
$\chi_i(k_z)$, $\gamma_i$, $\lambda_i$ to be independent of the in-plane direction ($x$ or $y$) such that the normal state topological quadrupolar semimetal has $C_4$ symmetry. To be explicit we will set $\lambda_i=\lambda$ and use it as our unit of energy with $\lambda=1$. We then set
$\chi_i(k_z)=\cos(k_z)/2$, and
$\gamma_i=\gamma=-3/4$.
With this choice of parametrization,  $h_{\rm SM}({\bf{k}})$ has fixed $k_z$  ``momentum slices'' with a non-vanishing quantized quadrupole moment for $\cos(k_z)<-1/2,$
and vanishing quadrupole for $\cos(k_z)>-1/2$.
As a consequence, the bulk bands are semimetallic with four-fold degenerate nodes at the locations $k_z^{(c)}$ where the quadrupole changes, i.e.,  $\cos(k_z^{(c)})=-1/2$~\cite{Lin2017}.
It is important to point out that any HOTSM with a well-defined nontrivial quadrupole moment and $C_4$ and mirror symmetries
should lead, in the presence of s-wave superconducting pairing, to results qualitatively similar to the one that we present below. The presence of the $C_4$
and mirror symmetries is important, as we will show, to be able to ``tune'' the hinge states via external magnetic fields.

\begin{table}[htb]
\begin{tabular}{|c|c|c|c|c|c|c|}
  \hline
  \hline
  Model & $\Lambda_i$ & $\mathcal{M}_x$ & $\mathcal{M}_y$ & $\mathcal{C}_4$ & Structure & HOTSC \\
  \hline
  \hline
  $h_{\rm SC,1}$ & $\sigma_1\kappa_2$ & $\tau_3 m_x$ & $\tau_3 m_y$ & - & intra-S & $\checkmark$\\
  \hline
  $h_{\rm SC,2}$ & $\sigma_2\kappa_1$ & $\tau_0 m_x$ & $\tau_3 m_y$ & - & intra-S & $\checkmark$\\
  \hline
  $h_{\rm SC,3}$ & $\sigma_2\kappa_0$ & $\tau_3 m_x$ & $\tau_3 m_y$ & - & inter-S & $\checkmark$\\
  \hline
  $h_{\rm SC,4}$ & $\sigma_2\kappa_3$ & $\tau_3 m_x$ & $\tau_0 m_y$ & - & inter-T & $\checkmark$\\
  \hline
  $h_{\rm SC,5}$ & $\sigma_0\kappa_2$ & $\tau_3 m_x$ & $\tau_3 m_y$ & $\tau_0 \hat{r}_4$ & inter-S & -- ($\checkmark^*$)\\
  \hline
  $h_{\rm SC,6}$ & $\sigma_3\kappa_2$ & $\tau_0 m_x$ & $\tau_0 m_y$ & $\tau_3 \hat{r}_4$ & intra-T & -- ($\checkmark^*$)\\
  \hline
\end{tabular}
\caption{The pairings for the six different models discussed in this work. Columns 2-4 show the representation of
the pairing term and of the symmetry operators  for each model.
$m_x=\sigma_1\kappa_3$, $m_y=\sigma_1\kappa_1$, $\hat{r}_4=\frac{i}{2}(\sigma_1+i\sigma_2)\kappa_2+\frac{1}{2}(\sigma_1-i\sigma_2)\kappa_0$.
Column 6 shows the pairing structure: inter/intra and S/T are short for inter/intra orbital and spin singlet/triplet, respectively.
Column 7 shows whether the model is a HOTSC:
models with "$^*$" in the parenthesis denotes existence of HOTSC in the presence of magnetic field.
}
\label{tab01}
\end{table}

The most general mean-field Hamiltonian describing a superconducting state for our system is given by $H_{\rm SC}=\sum_{\kk}\Psi_\kk^\dagger h_{\rm SC}(\kk)\Psi_\kk$
where $\Psi_\kk^T=\big(\psi^{\phantom{\dagger}}_{\kk},\,\psi^{\dagger}_{-\kk}\big)$ is the spinor in Nambu space
and $\hsc(\kk)=\tau_3 (h_{SM}(\kk)-\mu) + \Delta^{(ij)}_0(\kk)\tau_2\sigma_i\kappa_j$, where $\mu$ is the chemical potential, $\Delta^{(ij)}_0(\kk)$ the superconducting pairing strength in the $(ij)$
spin orbital channel, and $\{\tau_i\}$
are the Pauli matrices in Nambu space.
Restricting the superconducting pairing to be s-wave, i.e., $\Delta^{(ij)}_0(\kk)={\rm const}=\Delta_0$, we obtain
\beq
 h_{\rm SC,i} = \tau_3 (\hsm(\kk)-\mu) - \Delta_0\tau_2\Lambda_i
\enq
where $\Lambda_i$ is a $4\times 4$ matrix, independent of $\kk$, that determines the structure of the superconducting pairing
in orbital and spin space.
The requirement that the pairing term satisfies Fermi-Dirac statistics implies that there are only six possible
pairing matrices $\Lambda_i$, listed in the second column of Table~\ref{tab01}, (see e.g., \cite{Sato}).
As a consequence, starting from $\hsm$, we can obtain six distinct s-wave superconducting states.

The normal state already has broken time-reversal symmetry ($T^2=-1$), when the superconducting pairing is added these superconductors belong to symmetry class $D$~\cite{Chiu2016}. We note that, our model has a pseudo-chiral symmetry but its presence is not required for our results. All the models have mirror symmetries  $\mathcal{M}_x$, $\mathcal{M}_y,$ and $\mathcal{M}_z,$  and therefore overall inversion symmetry $\mathcal{I}=\mathcal{M}_x \mathcal{M}_y \mathcal{M}_z.$
The representation matrices for the $\mathcal{M}_x$ and $\mathcal{M}_y$  mirror symmetries are shown in Table~\ref{tab01}, and the matrix for $\mathcal{M}_z$ is the identity matrix. Models $h_{\rm SC,5}$ $h_{\rm SC,6}$, retain $C_4$ symmetry in the superconducting state with representation matrices given in Table~\ref{tab01}.

     Let us consider the quasi-particle spectra in the weak pairing limit $\Delta_0<\mu$. Explicitly, we choose $\mu=1, \Delta_0=0.5$ and find that models $h_{\rm SC,1},h_{\rm SC,3},h_{\rm SC,6}$ are fully gapped in the bulk by the pairing term while $h_{\rm SC,2}, h_{\rm SC,4}, h_{\rm SC,5}$ are gapless (see SM~\cite{sm}). Model $h_{\rm SC,2}$ ($h_{\rm SC,4}$) has gapless, bulk, quasi-particle states forming  a nodal line/loop in $k_x-k_z$ ($k_y-k_z$) plane, and model  $h_5$ has 4 nodes in the $k_x-k_y$ plane at $k_z$ values away from any high-symmetry points. We see that models $h_{\rm SC,1}$ and $h_{\rm SC,3}$ have the same symmetry, the same representation for  $\mx$ and $\my$, and have the same bulk spectra. Thus, they will have the same essential properties for our study, and henceforth we only consider $h_{\rm SC,1}.$ Additionally, models $h_{\rm SC,2}$ and $h_{\rm SC,4}$ are related by a unitary transformation and a $C_4$ rotation around the $z$-axis so we will only consider $h_{\rm SC,2}$ from now on. While we have focused on particular values of $\mu$ and $\Delta_0,$ the results do not change qualitatively as long as $\Delta_0<\mu$, as is appropriate for the weak-pairing limit.

\begin{figure}[htb]
\includegraphics[width=0.45\textwidth]{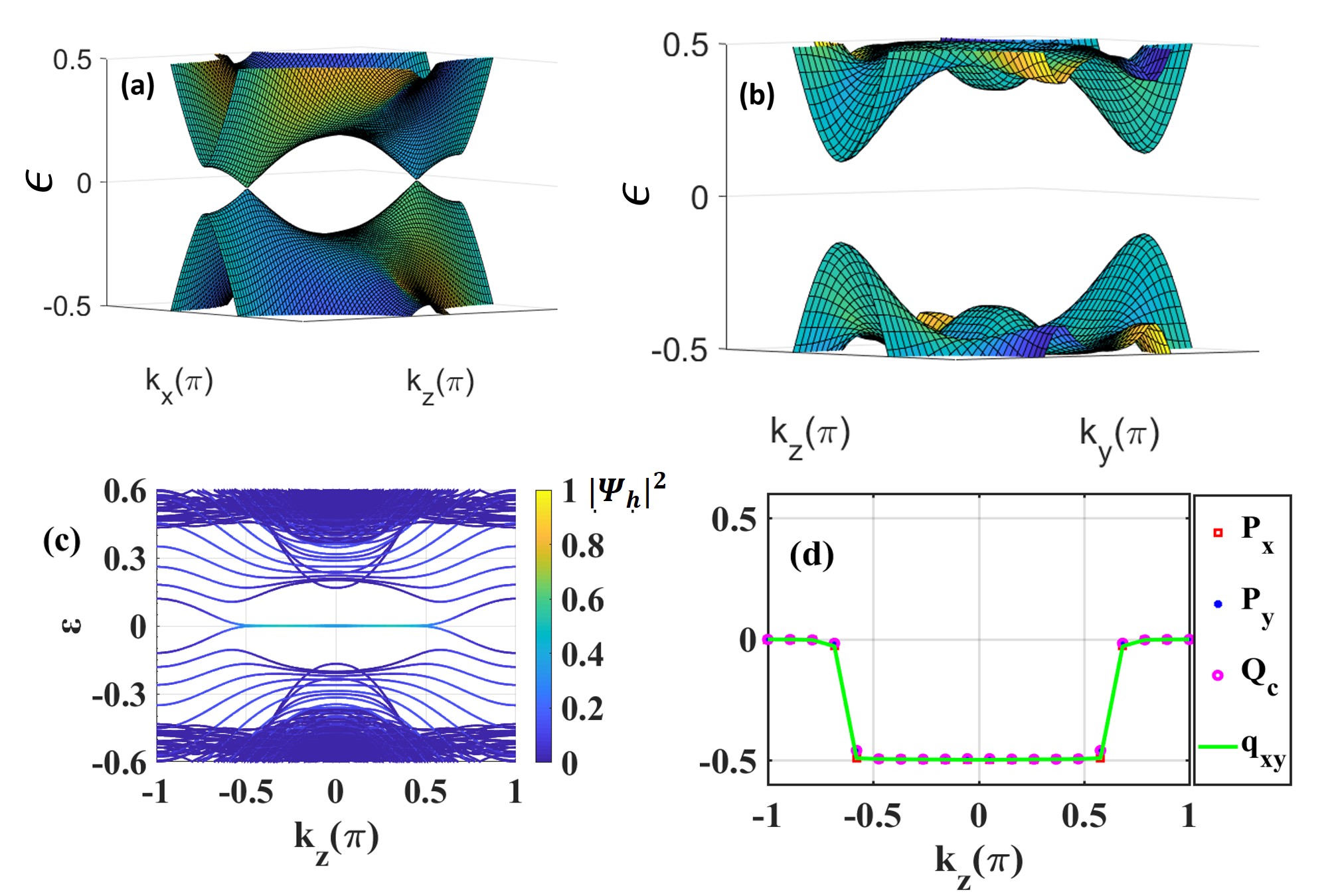}

\caption{The surface states of $h_{\rm SC,1}$ in (a) $k_x-k_z$ and (b) $k_y-k_z$planes. (c) shows the Majorana hinges arc states in $k_z$ direction.
(d) $P_x$, $P_y$, $Q_c$ and $q_{xy}$ versus $k_z$.}
\label{fig:h1}
\end{figure}

We start by considering model $h_{\rm SC,1}$. The pairing breaks $C_4$ symmetry, while retaining mirror symmetries, so the bulk are completely gapped\cite{Benalcazar2017-1} (see supplementary material (SM)~\cite{sm} for more detail). However, we find that the surfaces $S_{y,\pm}$, perpendicular to the $y$-axis, exhibit two gapless nodes per surface, Fig.~\ref{fig:h1}~(a),
whereas the ones perpendicular to the $x$-axis, $S_{x,\pm},$ are completely gapped, Fig.~\ref{fig:h1}~(b).
We then obtain the bands for the hinges and find,  as shown in Fig.~\ref{fig:h1}~(c), that dispersionless hinge Majorana states
are present at the four corners of the $xy$-plane
for values of $k_z$ between the two gapless nodes on the $S_y$ surface.
In Fig.~\ref{fig:h1}~(c) and all the other plots showing the spectrum of hinge states the color used for each point of the spectrum
denotes the magnitude of the square of the corresponding eigenstate at the hinges $|\Psi_h|^2$.
These states are reminiscent of the flat bands that appear between the nodes, and at the edges, of a 2D Dirac semimetal.

The structure of the spectrum and symmetry is similar to the one
of the quadrupolar semimetal obtained in Ref.~\onlinecite{Lin2017}
when $\gamma_i(k_z)\equiv \gamma_i +\chi(k_z)$ ($i=(x,y))$ follow what in Ref.~\onlinecite{Lin2017} is referred as path 2,
and suggests that the presence of the boundary states, the hinge
states in particular, might be due to a second-order topological invariant.
To confirm it, we calculated (for details see SM~\cite{sm}) the superconducting analog of the quadrupole moment $q_{xy}(k_z)$ for each $k_z$ slice.
We have $q_{xy}(k_z)\equiv (P_x(k_z) + P_y(k_z) - Q_c(k_z))\mod 1$~\cite{Benalcazar2017-1}
where $P_x(k_z), P_y(k_z)$  are the surface polarizations in the $x$ and $y$ directions, respectively,
and $Q_c(k_z)$ is the corner charge which takes values $1/2, (0)\mod\,1$, if hinge states are present (absent).

Fig.~\ref{fig:h1}~(d) shows that $P_x(k_z)$, $P_y(k_z)$ are quantized, and that they take the non-trivial value -1/2 for
$k_z$ in the interval between the two gapless nodes of the surfaces states on $S_y$, the same range of $k_z$ for which we have hinge states.
As a consequence we find
that for values of $k_z$ between the two surface nodes,  $q_{xy}(k_z)$  is also non-trivial and therefore that the hinge states in model $h_{\rm SC,1}$ are topologically protected and can be captured by a second-order, quadrupole-like, topological invariant. Interestingly, we find that the presence of the gapless nodes on the surface, and of the hinge states, is not affected by perturbations that break both mirror symmetries, and thus the hinge states are perturbatively stable when the BdG particle-hole symmetry is maintained even when the mirror symmetry is broken (see SM~\cite{sm}).

Due to the presence of Dirac nodes in the band structure of the surface states, and finite value of the quadrupole-like invariant we term
superconductors like the one described by model $h_{\rm SC,1}$ (and $h_{\rm SC,3}$) {\em second order Dirac superconductors}.
This is one of the main results of this work.

For models $h_{\rm SC,{2,4}}$ the gapless nature of the bulk and surface states leads to obscured hinge modes (see Fig. 2 in SM \cite{sm}).

\begin{figure}[htb]
\includegraphics[width=0.45\textwidth]{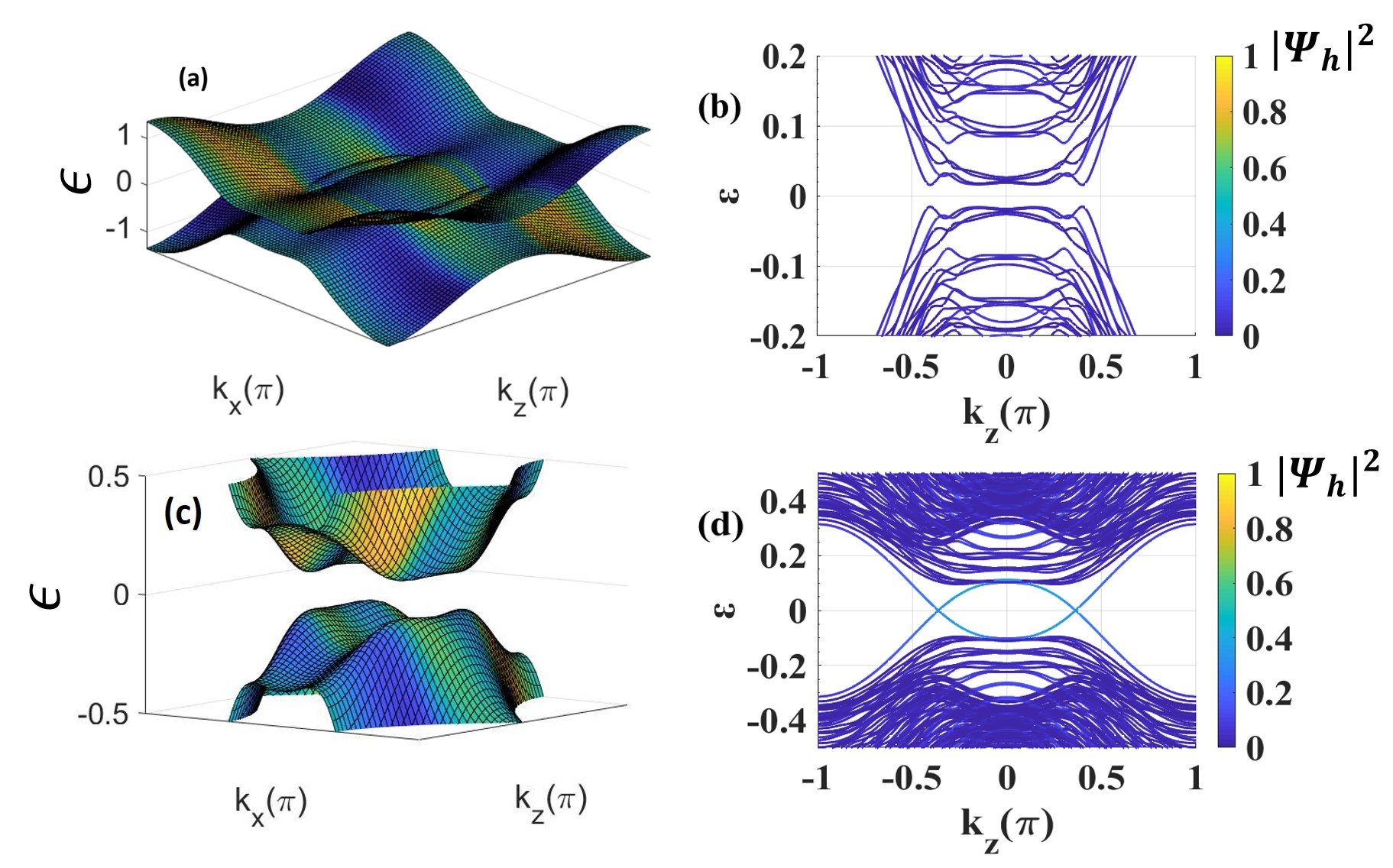}
  \caption{Model $h_{\rm SC,5}$: Surface states (a) in $k_x-k_z$ plane, and hinge states (b) along $k_z$, when $J_z=0$,
 (c) and (d), same as (a) and (b), respectively, for the case when $J_z=0.4$.
  }
 \label{fig:h5}
\end{figure}
\begin{figure}[htb]
\includegraphics[width=0.45\textwidth]{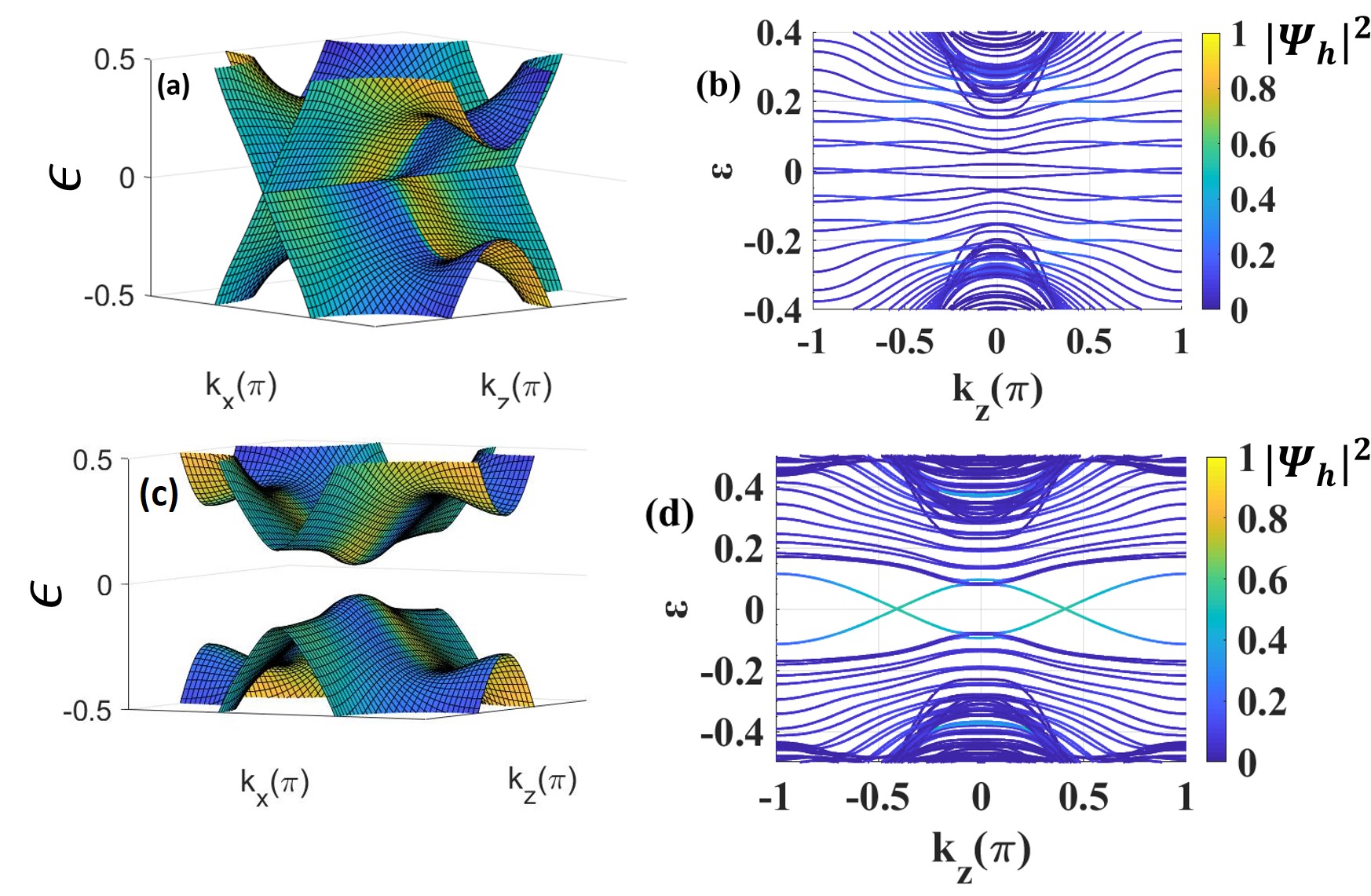}
   \caption{Model $h_{\rm SC,6}$: Surface states (a) in $k_x-k_z$ plane, and hinge states (b) along $k_z$, when $J_z=0$,
 (c) and (d), same as (a) and (b), respectively, for the case when $J_z=0.4$.
  }
 \label{fig:h6}
\end{figure}

Models $h_{\rm SC,5}$ and $h_{\rm SC,6}$ differ from the previous models in that, in addition to the mirror symmetries $\mx$ and $\my$,
they retain the $C_4$ symmetry of the original semimetal.

As mentioned,  model $h_{\rm SC,5}$ has nodal points in the bulk quasi-particle spectrum, while $h_{\rm SC,6}$ is fully gapped in the bulk (see SM~\cite{sm}).
We find that both models have gapless surface states but of different nature. Due to the $C_4$ symmetry, we only show the results for the surface states  $S_{y,\pm}$.
The surface bands of model $h_{\rm SC,5}$ exhibit two nodal loops, see Fig.~\ref{fig:h5}~(a),  whereas the surface bands of model  $h_{\rm SC,6}$
exhibit a nodal line at $k_x=0$ for the $S_y$ surface states ($k_y=0$ for the $S_x$ surface states), Fig.~\ref{fig:h6}~(a).
Figure~\ref{fig:h5}~(b) and Fig.~\ref{fig:h6}~(b) show the bands for the hinges for models $h_{\rm SC,5}$ and $h_{\rm SC,6}$ and we see that there are no interesting localized hinge modes present.

\begin{figure}[htb]
 \includegraphics[width=0.99\columnwidth]{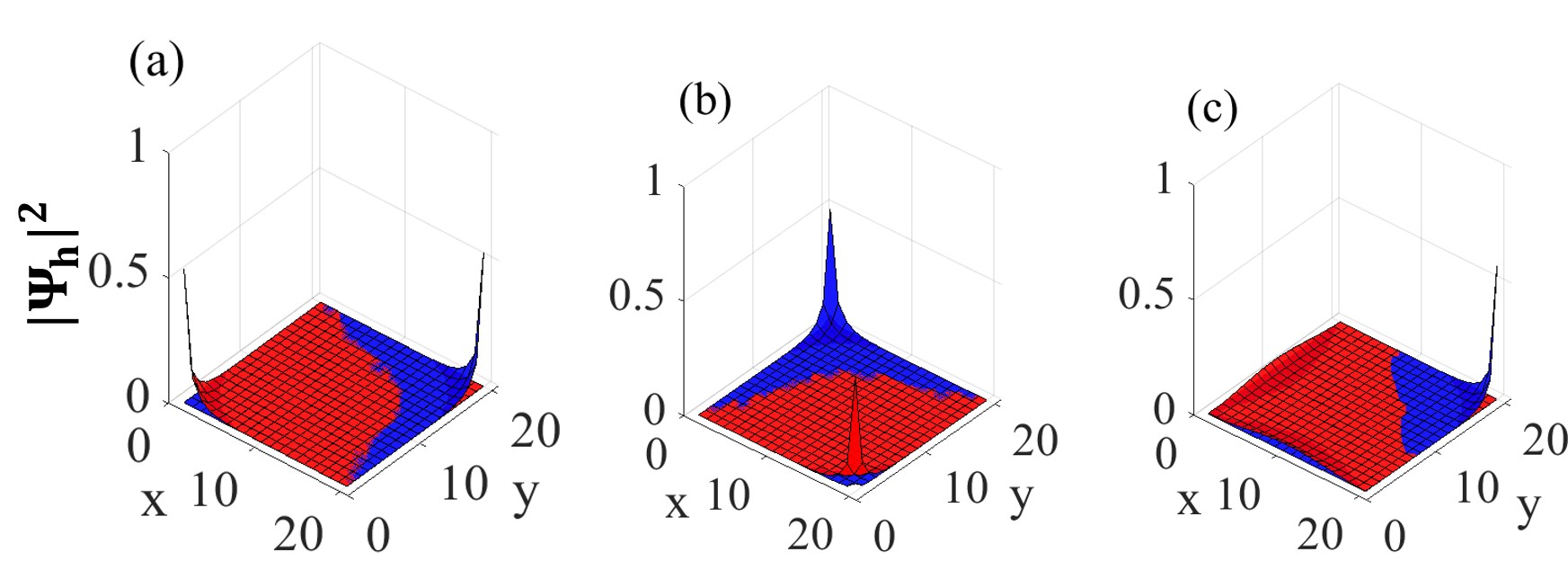}
   \caption{$|\Psi_h(x,y)|^2$ for model $h_{\rm SC,6}$ in the presence of a Zeeman term.
     We have two pairs of degenerate states. For each pair $|\Psi_h(x,y))|^2$ is the same. In the figure
     we show in blue and red the two different $|\Psi_h(x,y))|^2$ profiles.
 (a) $J_z=0.4$, $k_z=0.41\pi$, value for which the helical hinge bands cross (Fig.~\ref{fig:h6}~(d)).
(b) $J_z=-0.4$, $k_z=0.41\pi$.
(c) $J_z=0.4$ plus $C_4\mathcal{M}_x$ breaking perturbation: $0.1\tau_3r_4^{-1} m_x$. 
 In (c) $k_z=0.19\pi$, for which we have the lowest positive energy.}
 \label{fig:cornerh6}
\end{figure}

However, an external magnetic field
can perturb these systems to generate hinge modes.
Let us apply a uniform magnetic field, or proximity-couple to a ferromagnet,
to generate a Zeeman term $H_z=J_z\tau_3\sigma_3\kappa_0$, where $J_z$ is directly proportional to the magnitude
of the external magnetic field (we ignore any orbital effects of the magnetic field). This term
qualitatively modifies the band structures of $h_{\rm SC,5}$ and $h_{\rm SC,6}$.
From Figs.~\ref{fig:h5}~(c),~\ref{fig:h6}~(c) we see that the presence
of the Zeeman term gaps out the the surface states completely (it also gaps out the bulk nodes of $h_{\rm SC,5}$). Furthermore, we see the appearance of
clear  hinge states within the gap of the surface states, as shown in Fig.~\ref{fig:h5}~(d) and Fig.~\ref{fig:h6}~(d). $H_z$ breaks $C_4$ symmetry and both mirror symmetries, but leaves the products  $C_4 \mathcal{M}_x$ (anti-diagonal mirror) and $C_4 \mathcal{M}_y$ (diagonal mirror) intact. Because of these symmetries, and contrary to the hinge states of the other models discussed in this work, the hinge states of $h_{\rm SC,5}$ and $h_{\rm SC,6}$ (with Zeeman) are: dispersive, non-chiral, and localized only at two of the four hinges related by $C_2$ symmetry.
In addition, we find that the pair of corners where the helical hinge states are localized  switches upon a change of sign of the Zeeman term (e.g., switching the direction of the external magnetic field), see Fig.~\ref{fig:cornerh6}. This phenomenon could be useful for the experimental detection of these systems.
We note that magnetic fields were also proposed to induce a HOTSC state in a completely different system with different properties in two dimensions \cite{Zhu2018}.

For $J_z>0$ ($J_z<0$) there  are two, counter-propagating modes at two of the four corners, and they are $\pm$ eigenstates of $C_4 \mathcal{M}_y$ ($C_4 \mathcal{M}_x$). The hinge modes are perturbatively stable even in the absence of the two $C_4\mathcal{M}_i$ symmetries as long as particle-hole symmetry is preserved.
However, the modes can be destroyed through bulk or surface phase transitions.
Because the hinge states are located at opposite sides of the anti-diagonal (diagonal), the breaking of the $C_4 \mathcal{M}_x$  ($C_4 \mathcal{M}_y$) symmetry leads to non-symmetric hinge states and, for a strong enough   $C_4 \mathcal{M}_x$  ($C_4 \mathcal{M}_y$) symmetry breaking perturbation, to the almost complete suppression of the hinge states at one corner and the enhancement of the hinge states at the opposite corner, as shown in Fig. \ref{fig:cornerh6}~(c).

In conclusion, we have identified a new class  of higher-order topological superconductors (HOTSCs): {\em second-order Dirac superconductors}.
HOTSCs in this class  have both topologically protected Dirac cones for states at the surface (d-1 boundary) and topologically protected dispersionless Majorana states at the hinges (d-2 boundary). The coexistence of gapless surface and hinge modes should lead to novel transport properties that
it would be interesting to compare to the ones of conventional topological superconductors\cite{fu2009probing,fu2009josephson,Foster2012,Foster2014,Ghorashi2017,Ali2018,Roy-Ghorashi,GhorashiFoster2019,Xie2015,Chiu2016,FanZhang1,FanZhang2}.
We have then identified 3D superconducting systems that can be driven into a higher-order topological state simply via inclusion of a Zeeman term, having dispersive helical Majorana hinge states that are located at only two (or one) of the corners.
We have shown that by varying the relative sign of the Zeeman term and of the mirror-diagonal symmetry-breaking perturbation, the position of the Majorana pairs can be tuned to be at a specific corner, leading to a highly tunable setup for realizing, {\em and} detecting, helical Majorana states.


\begin{acknowledgments}
S.A.A.G., X.H. and E.R. acknowledge support from ARO (Grant No. W911NF-18-1-0290), NSF (Grant No. DMR-1455233) and ONR (Grant No. ONR-N00014-16-1-3158). E.R. thanks the Aspen Center for Physics, which is supported by National Science Foundation grant PHY-1607611, for its hospitality during the early stages of the work.
The numerical calculations have been performed on computing facilities at William \& Mary which were provided by contributions from the NSF, the Commonwealth of Virginia Equipment Trust Fund, and ONR. TLH thanks the US
National Science Foundation under grant DMR 1351895-
CAR, and the MRSEC program under NSF Award Number
DMR-1720633 (SuperSEED) for support.
\end{acknowledgments}


\bibliography{hosc}



\newpage \clearpage

\onecolumngrid

\begin{center}
	{\large
	Second-order Dirac superconductors and magnetic field induced Majorana hinge modes
	\vspace{4pt}
	\\
	SUPPLEMENTAL MATERIAL
	}
\end{center}

\section{I.\, Bulk spectrum and continuum analysis}
\begin{figure}[H]
\includegraphics[width=0.9\textwidth]{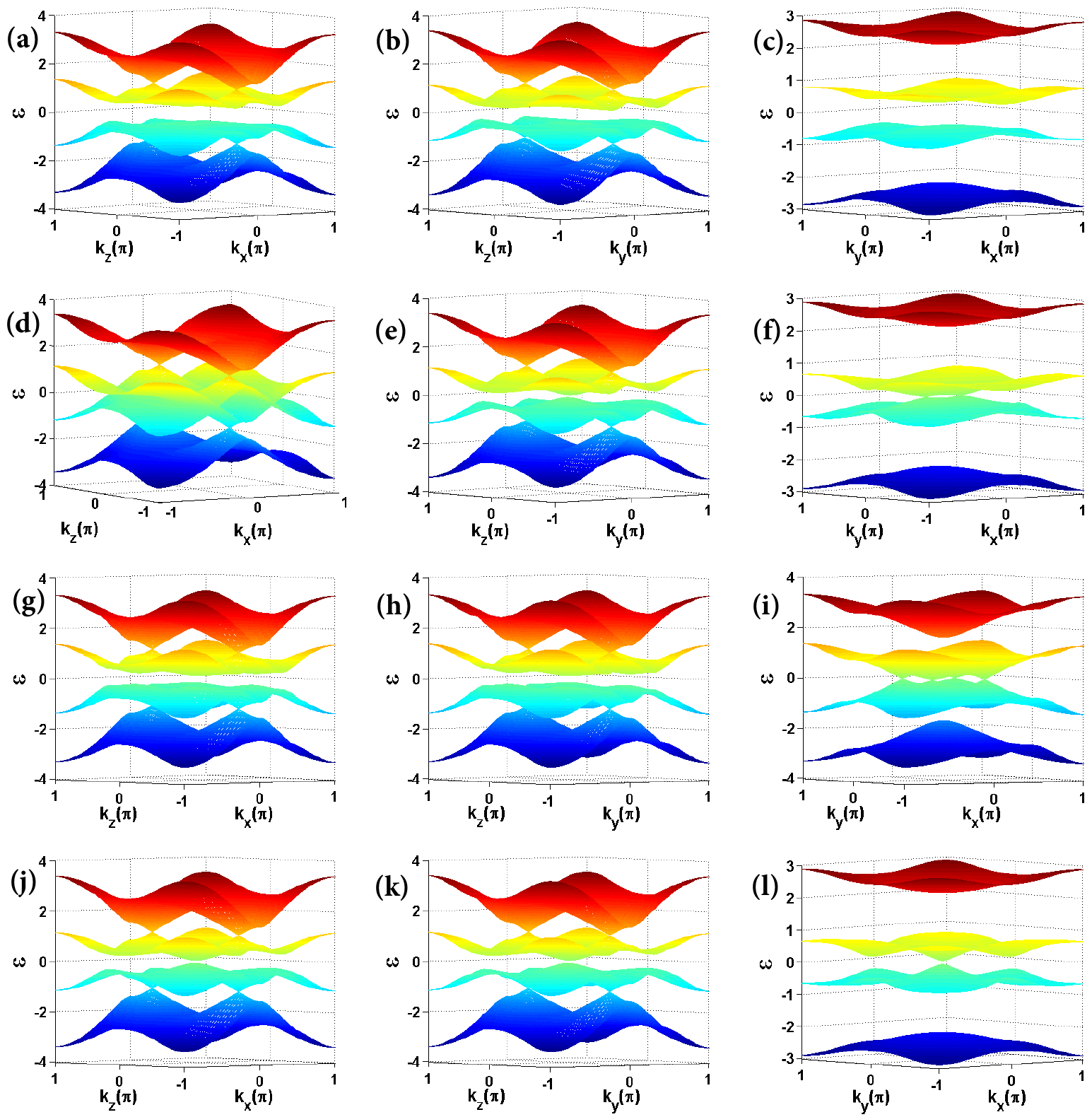}
\caption{Bulk spectrum for models $h_{\rm SC,1}, h_{\rm SC,2}, h_{\rm SC,5}$ and $h_{SC,6}$ shown in the first, second, third and fourth rows, respectively. Each row, from left to right, represents $k_x-k_z$ ($k_y=0$), $k_y-k_z$ ($k_x=0$) and $k_x-k_y$ ($k_z=0$ for all model except $h_{SC,5}$ where $k_z=0.4$ is used) plane.}
\label{fig:bulk}
\end{figure}

We can analyze the bulk quasi-particle spectra by considering the effects of $\Delta_0$ perturbatively on the normal state semimetal by using a continuum ${\bf k}\cdot{\bf p}$ expansion around the nodes at the two $k_{z}^{(c)}.$ The $16\times 16$ continuum Hamiltonian is
\begin{eqnarray}
   H_{{\bf k}\cdot{\bf p},i}&=&-\tau_3\pi_0\sigma_2(vk_x\kappa_3+vk_y\kappa_1)+v_zk_z(\tau_3\pi_3\sigma_2\kappa_2\nonumber\\&-&\tau_3\pi_3\sigma_1\kappa_0)
   -\mu\tau_3\pi_0\sigma_0\kappa_0-\Delta_0\tau_2\pi_1\Lambda_i
    \end{eqnarray}
\noindent where $\pi_i$ are Pauli matrices in the valley/node degree of freedom, and $v, v_z$ are velocities $(\hbar=1)$. In the nodal limit $\mu\to 0$, one can determine the nature of the bulk spectra by calculating how the pairing term commutes/anti-commutes with the kinetic terms.\\
In Fig~\ref{fig:bulk}, we present bulk band structures for all the models discussed in this work using the lattice model of Eq. (2) in the manuscript.\\

\newpage

\section{Model $h_{sc,2}$}

\begin{figure}[htb]
\includegraphics[width=0.35\textwidth]{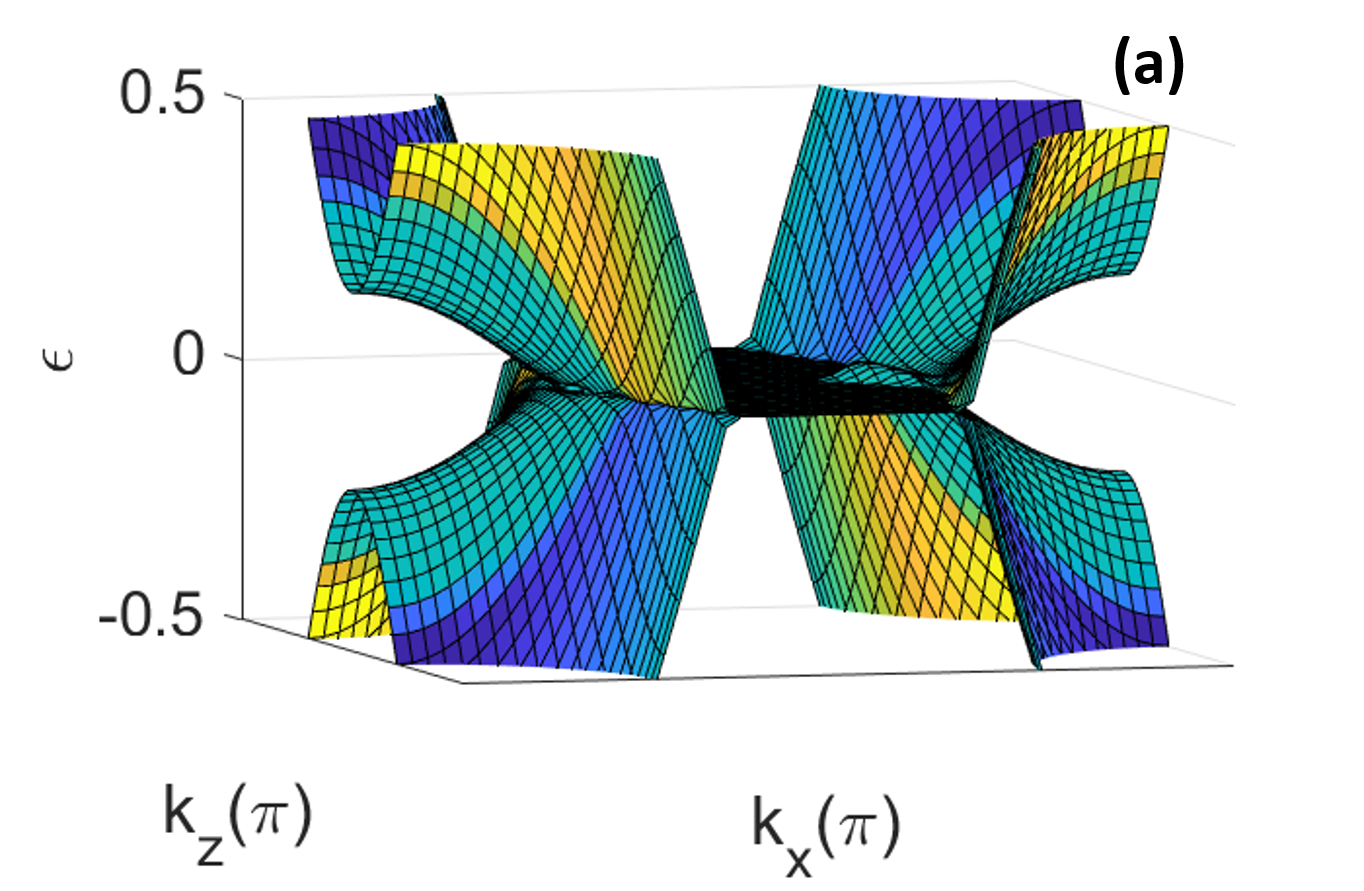}\includegraphics[width=0.35\textwidth]{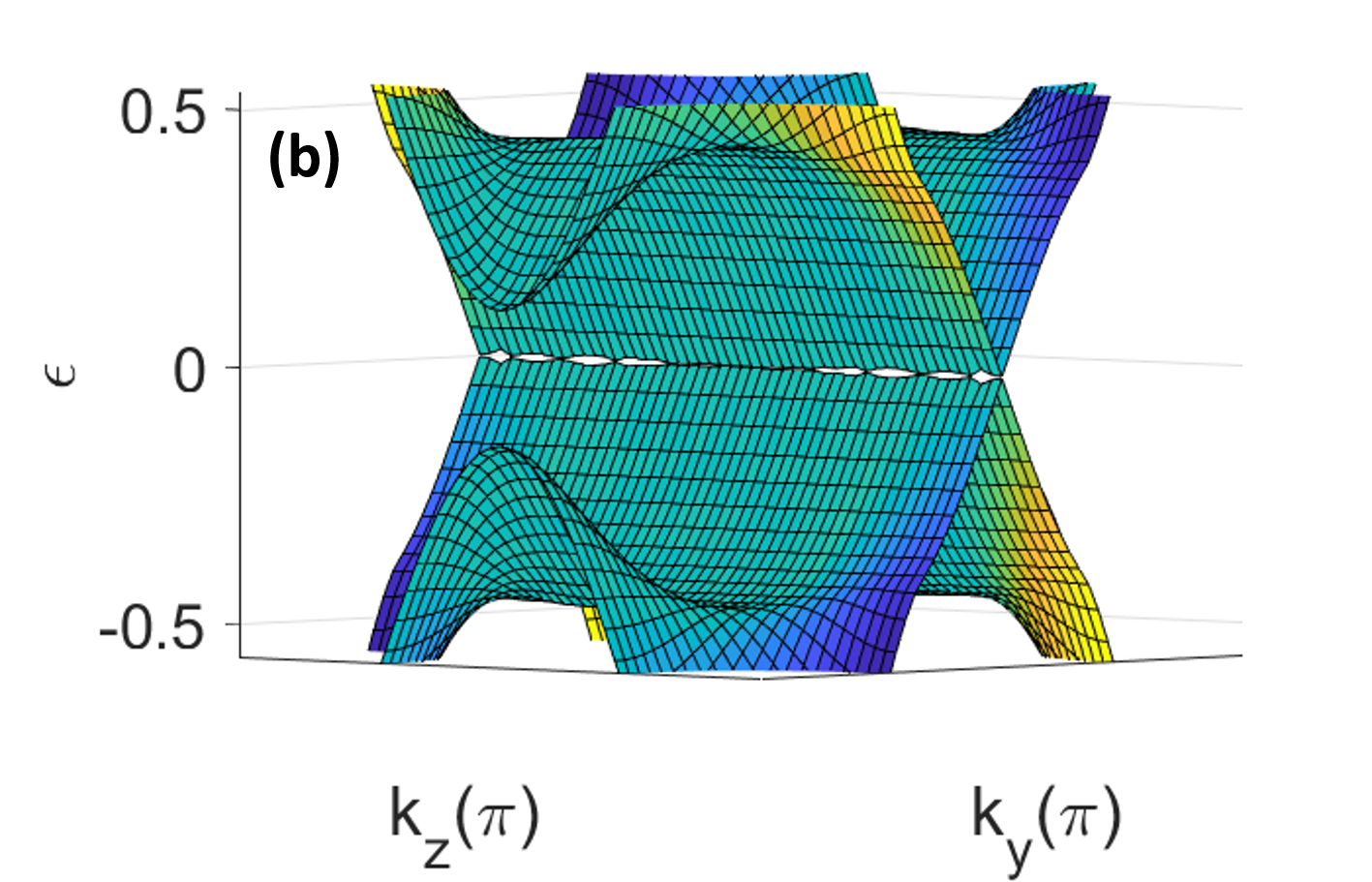}
\includegraphics[width=0.35\textwidth]{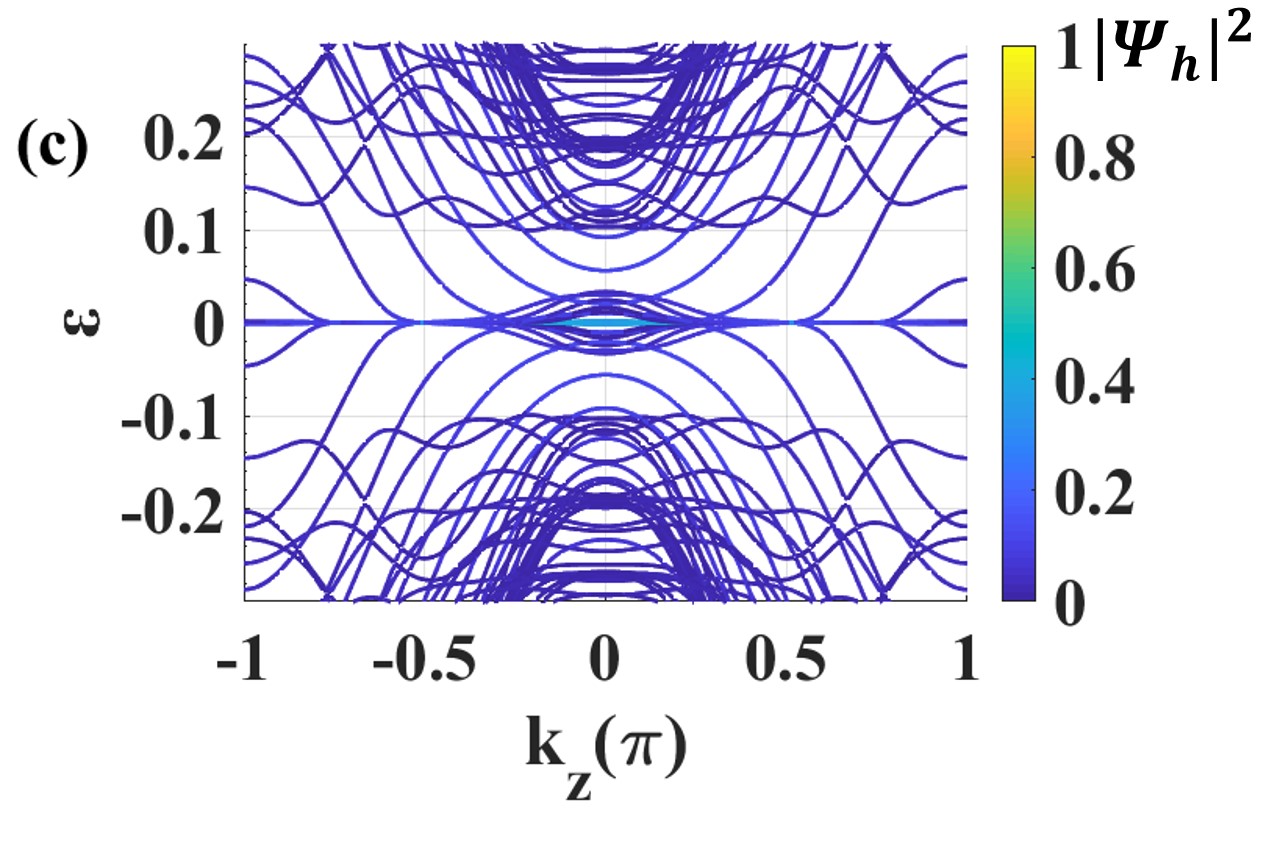}
\caption{The surface states of $h_{\rm SC,2}$ in (a) $k_x-k_z$ and (b) $k_y-k_z$planes. (c) shows the Majorana hinges arc states in $k_z$ direction.}
\label{fig:h2}
\end{figure}

As mentioned in the manuscript the bulk states of model $h_{\rm SC,2}.$ have nodal loops, (see Fig~\ref{fig:bulk}), and the surface states are gapless, as shown in Figs.~\ref{fig:h2}~(a),~(b).
While we do find a small region of localized hinges states in the middle of the spectrum (see Fig.~\ref{fig:h2}~(c)), the gapless bulk and surface states obscure them. It would be difficult to spectroscopically isolate the hinge modes of this model for any practical or experimental purpose. The same is true of model $h_{\rm SC,4}.$ Thus, as we noted in the manuscript, while there may be some interesting features of the gapless bulk and surface states of these models left to explore, we leave the study of these models for future works given that in this work our primary interest is in hinge-mode phenomenology.

\section{The effect of symmetry-breaking perturbations}

In this section we show that the hinges states corresponding to second-order topological superconductors of $h_{\rm SC,1}$ ( or $h_{\rm SC,3}$) are robust against mirror-symmetry breaking perturbations, and that the effect of the Zeeman term $J_z$ on the obscured hinges states of models $h_{\rm SC,2}$ ( or $h_{\rm SC,4}$).

\begin{figure}[htb]
\includegraphics[width=0.4\textwidth]{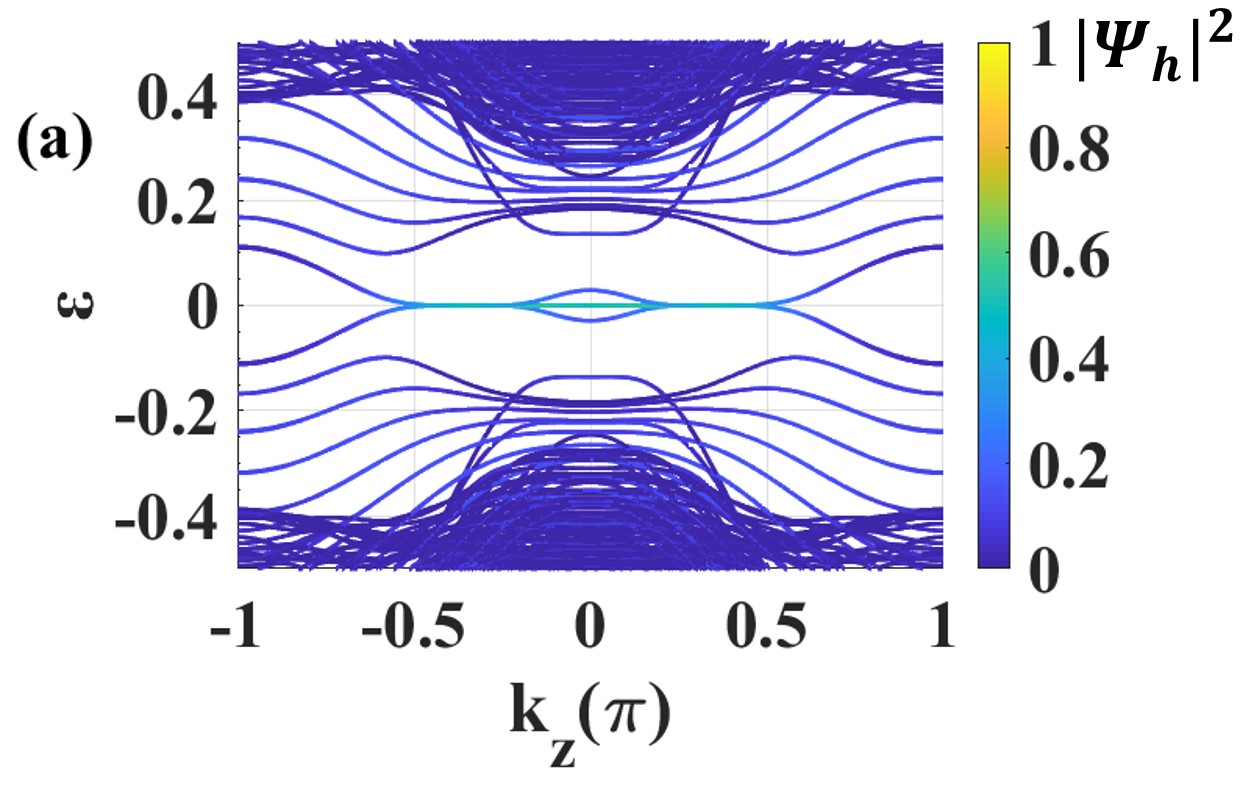}\includegraphics[width=0.4\textwidth]{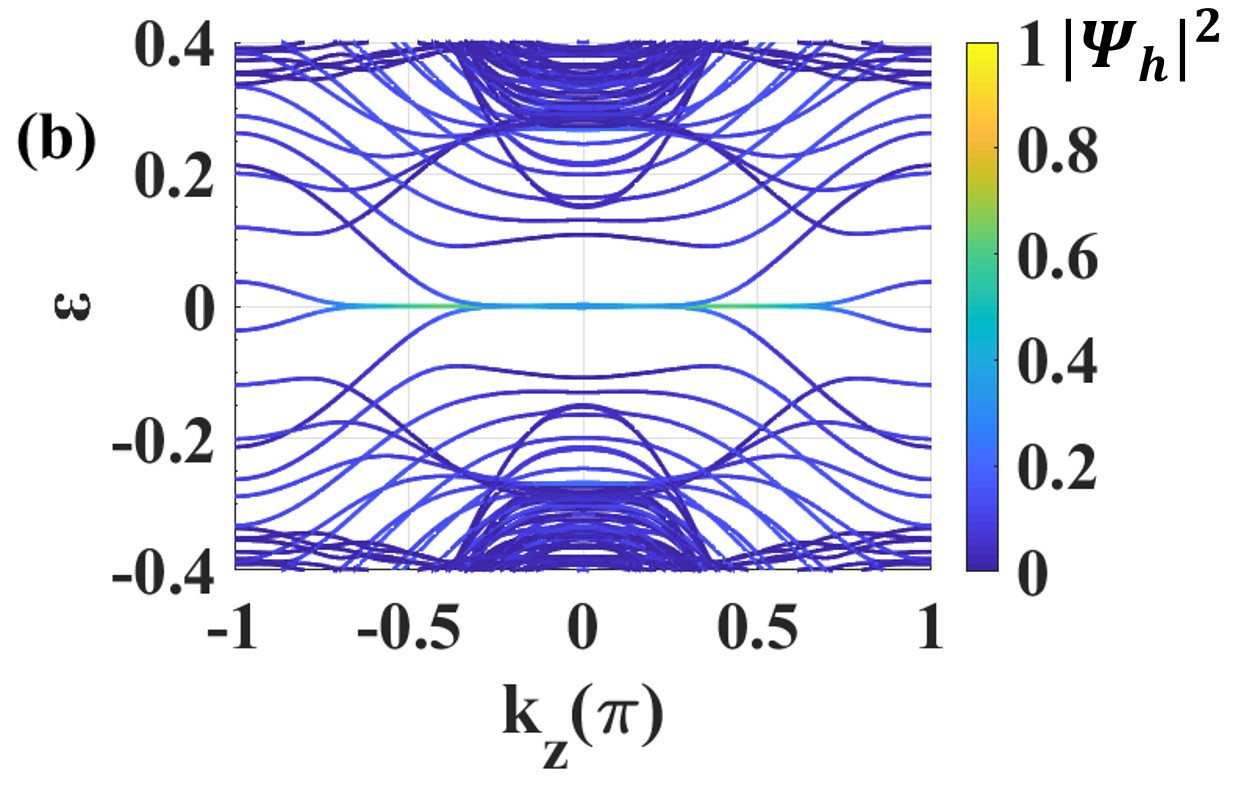}
\caption{The hinge states along $k_z$-direction for $h_{\rm SC,1}$ in the presence of (a) $\mathcal{M}_x$ and (b) $\mathcal{M}_y$ breaking perturbations. $\mu=1,\Delta_0=0.5$ and perturbation strength of $0.1$ is used.}
\label{fig:h1SM}
\end{figure}

Figures \ref{fig:h1SM}(a) and (b) show the spectrum of hinge states for model $h_{\rm SC,1}$ in the presence of weak $\mathcal{M}_x$ and $\mathcal{M}_y$ breaking perturbations. As is evident hinge states are robust against weak symmetry-breaking perturbations and at most can partially remove degeneracies.

\begin{figure}[htb]
\includegraphics[width=0.4\textwidth]{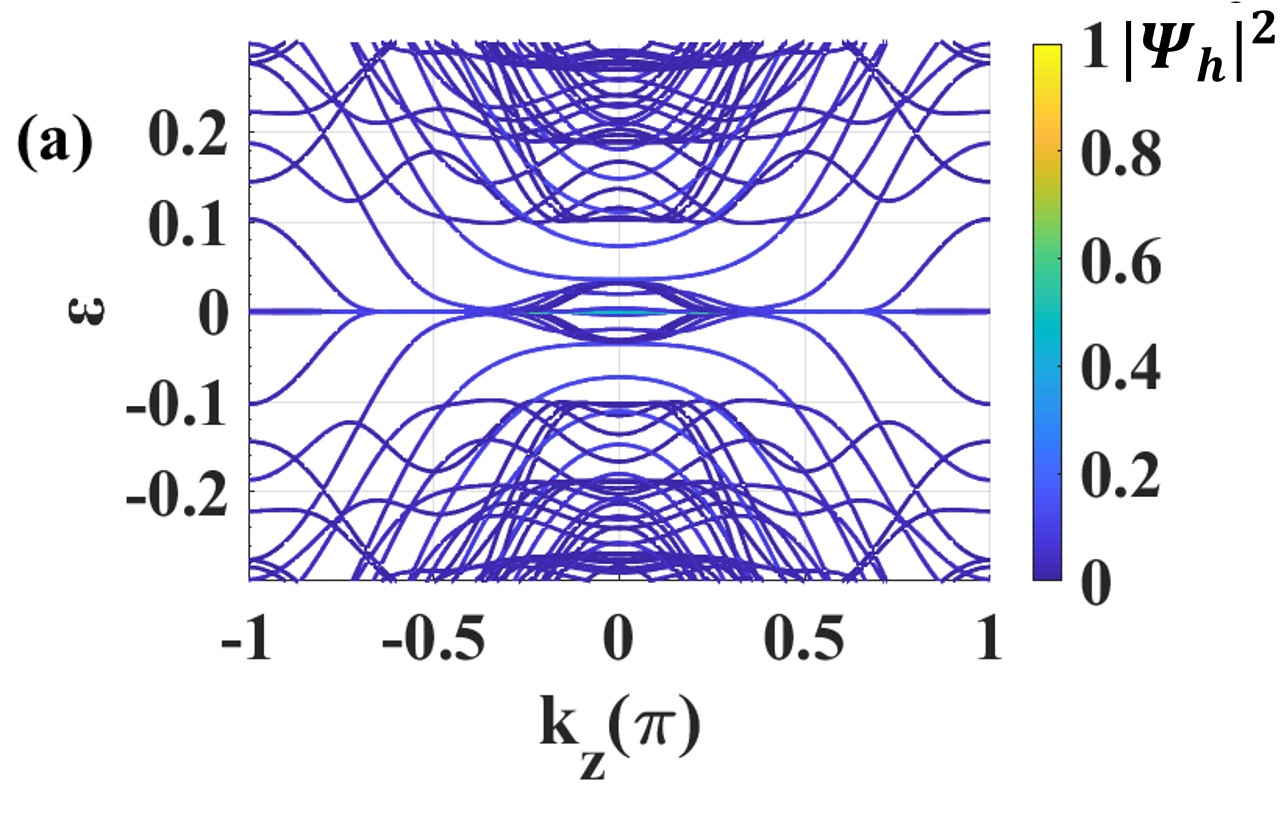}\includegraphics[width=0.4\textwidth]{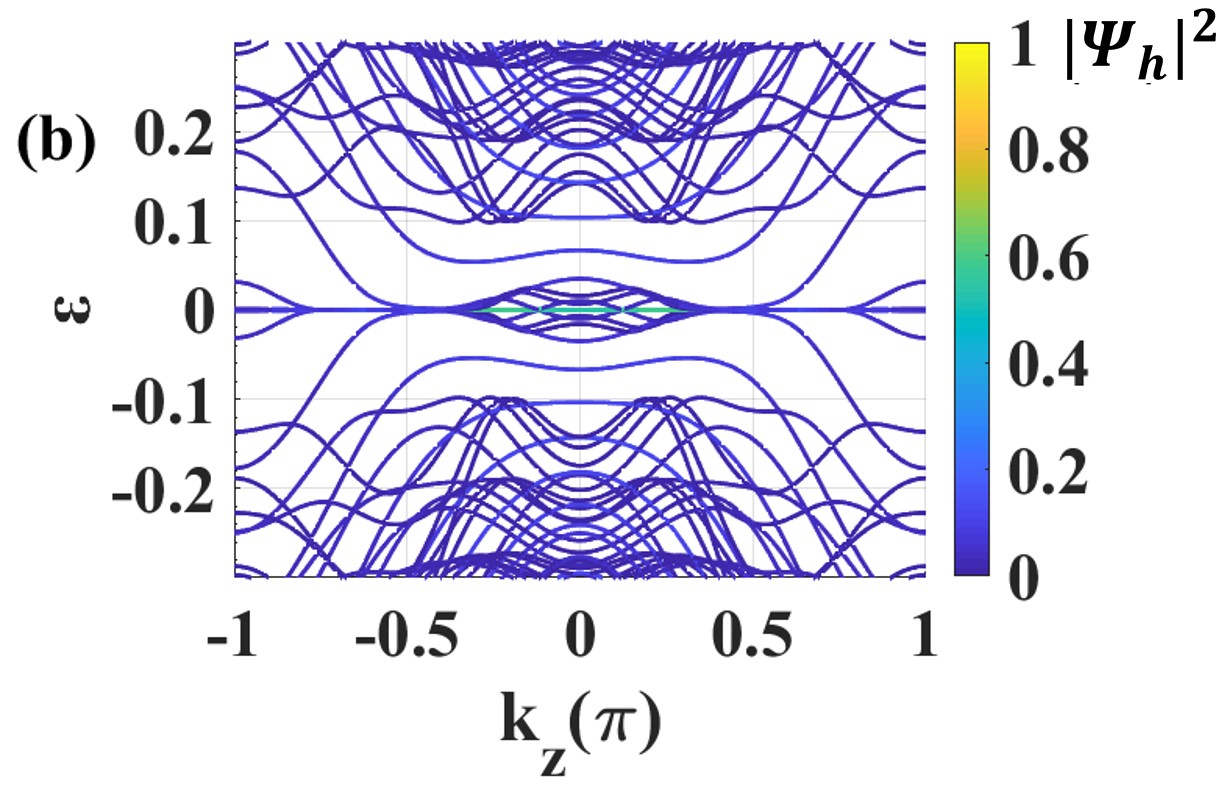}
\caption{The hinge states along $k_z$-direction for $h_{\rm SC,2}$ in the presence of the Zeeman terms with (a) $J_z=0.4$ and (b) $J_z=0.6$. $\mu=1,\Delta_0=0.5$ is used. }
\label{fig:h2SM}
\end{figure}

Fig. \ref{fig:h2SM}, shows the hinge spectrum for model $h_{\rm SC,2}$ in the presence of a magnetic field along the $z$ direction, $J_z$, for two
different values of $J_z$. We observe that the Zeeman term can only gradually remove some of the gapless points. We could not find any weak perturbation that can fully gap out the hinges states while preserving particle-hole symmetry.

\section{The calculation of the quadrupole-like invariant for the second-order Dirac superconductors}

For each superconducting Hamiltonian with two orbitals and one spin degree of freedom
$H_{\rm SC,i}=\Psi_\kk^\dagger(\tau_3 \hsm(\kk) - \Delta_0\tau_2\Lambda_i)\Psi_\kk$,  we can substitute the original basis $\Psi_{\kk}=(c_{\kk\up},d_{\kk\up},c_{\kk\down},d_{\kk\down},c_{-\kk\up}^\dagger,d_{-\kk\up}^\dagger,c_{-\kk\down}^\dagger,d_{-\kk\down}^\dagger)^T$ with a new basis $\tilde{\Psi}_{\kk}=(c_{\kk\up},d_{\kk\up},c_{\kk\down},d_{\kk\down},\tilde{c}_{\kk\up},\tilde{d}_{\kk\up},\tilde{c}_{\kk\down},\tilde{d}_{\kk\down})^T$, where $\tilde{c}$ and $\tilde{d}$ are two additional electron orbitals. Now we have a Hamitonian with the pure electron basis
\beq
\tilde{H}_i=\sum_{\bf k}\tilde{\Psi}_{\kk}^{\dagger}(\tau_3 \hsm(\kk) - \Delta_0\tau_2\Lambda_i)\tilde{\Psi}_{\kk}.
\enq
The formulation of the original BdG Hamiltonian in terms of purely electronic degrees of freedom makes direct
the connection between the quadrupole moment of non-superconducting systems and the equivalent ``quadrupole moment''
that can be used to identify higher order topological superconducting states.
%
%

The calculation of the polarization and quadrupole for model $h_{\rm SC, 1 (3)}$ is carried out using the method presented in Ref.[7].
As stated in the main text we have
$q_{xy}(k_z)\equiv (P_x(k_z) + P_y(k_z) - Q_c(k_z))\mod 1$,
where $q_{xy}$ is the quadrupole moment, $P_x(k_z), P_y(k_z)$  are the surface polarizations in the $x$ and $y$ directions, respectively,
and $Q_c(k_z)$ is the corner charge which takes values $1/2, (0)\mod\,1$, if hinge states are present (absent).
To obtain the edge-localized polarization $P_x(k_z)$ for a fixed $k_z$, we place the Hamiltonian
$\tilde{H}_{1(3)}(k_z)$ in a geometry which is open in the $y$ direction and periodic in the $x$ direction.
We have $P_x=\sum_{R_y=N_y-N_l+1}^{N_y}p_x(R_y)$, where $R_y$ is a site along $y$ in the discretized model,
 $N_y$ is the size of the system along the $y$ direction (in number of sites of the discretized model),
$N_l$ is the finite thickness of the edge layer used in the calculation, and
(see Eqs.(5.8) and (5.9) of Ref.[7]):
\begin{equation}
p_x(R_y)=\sum_j\rho^j(R_y)\nu_x^j;
\end{equation}
where
\begin{equation}
\rho^{j,R_x}(R_y)=\frac{1}{N_x}\sum_{k_x,\alpha}|[u_{k_x}^n]^{R_y,\alpha}[\nu_{k_x}^j]^n|^2,
\end{equation}
In the equations above
$j$ and $\nu_x^j$ are the index and eigenvalue for the Wilson loop along $k_x$, respectively,
$R_x$ and $R_y$ are the position in $x$ and $y$ directions. $N_x$ is the size in $x$ direction,
$[\nu_{k_x}^j]^n$ represent the $n-$th component of the $j-$th Wannier function, and
$[u_{k_x}^n]^{R_y,\alpha}$ is the component of the $n-$th Bloch wave function
at $R_y$ with orbital index $\alpha$.
$P_y$ is obtained in a similar way via the substitutions $x\to y$, $y\to x$.
We calculate the corner charge using open boundary conditions both in the $x$ and
$y$ direction:
\beq
Q_c=\sum_{R_x=N_x-N_l+1}^{N_x}\;\sum_{R_y=N_y-N_l+1}^{N_y}\rho({\bf R}),
\enq
where $\rho({\bf R})$ is the charge density due to the occupied states.

In our calculations, we used $N_x=50$, $N_y=50$, $N_z=19$ and $N_l=10$. For the bulk Hamiltonian, four bands are occupied.
The values of the other parameters are the ones stated in the main text.
When calculating the quadrupole moment for model  $h_{\rm SC,1}$ we add a small perturbation of the form
\beq
\delta\tilde{H}_1=0.005\tau_0\Gamma_0+0.010(\tau_3\sigma_1\kappa_1+\tau_3\sigma_0\kappa_3).
\enq
in order to fix the sign of $P_x$, $P_y$ and $Q_c$ (i.e. to remove the ambiguity due to the fact that these quantities are well definide only up to $\mod 1$).

\end{document}